# Quantum-Secured DSP-Lite Data Transmission Architectures for AI-Driven Data Centres

Xitao Ji[1,#], Wenjie He[2,#], Junda Chen[1], Mingming Zhang[1], Yuqi Li[1], Ziwen Zhou[1], Zhuoxuan Song[1], Hao Wu[1], Siqi Yan[1,4,*], Kejin Wei[2,*], Zhenrong Zhang[2], Shuang Wang[3] and Ming Tang[1,4]

[1]School of Optical and Electronic Information, Huazhong University of Science and Technology, Wuhan 430074, China

[2]Guangxi Key Laboratory for Relativistic Astrophysics, School of Physical Science and Technology, Guangxi University, Nanning 530004, China

3 CAS Key Laboratory of Quantum Information, Hefei National Laboratory, Hefei 230088, China, University of Science and Technology of China, Hefei, Anhui 230026, China

4 Hubei Optical Fundamental Research Centre, Wuhan 430074, China

[*]Email:siqya@hust.edu.cn and kjwei@gxu.edu.cn

[#]These authors equally contributed to this work.

# Abstract

Artificial intelligence-driven (AI-driven) data centres, which require high-performance, scalable, energy-efficient, and secure infrastructure, have led to unprecedented data traffic demands. These demands involve low latency, high bandwidth connections, low power consumption, and data confidentiality. However, conventional optical interconnect solutions, such as intensity-modulated direct detection and traditional coherent systems, cannot address these requirements simultaneously. In particular, conventional encryption protocols that rely on complex algorithms are increasingly vulnerable to the rapid advancement of quantum computing. Here, we propose and demonstrate a quantum-secured digital signal processing-lite (DSP-Lite) data transmission architecture that meets all the stringent requirements for AI-driven data centre optical interconnects (AI-DCIs) scenarios. By integrating a self-homodyne coherent (SHC) system and quantum key distribution (QKD) through the multicore-fibre-based space division multiplexing (SDM) technology, our scheme enables secure, high-capacity, and energy-efficient data transmission while ensuring resilience against quantum computing threats. In our demonstration, we achieved an expandable transmission capacity of 2 Tbit per second (Tb/s) and a quantum secret key rate (SKR) of 229.2 kb/s, with a quantum bit error rate (QBER) of approximately 1.27% and with ultralow power consumption. Our work paves the way for constructing secure, scalable, and cost-efficient data transmission frameworks, thus enabling the next generation of intelligent, leak-proof optical interconnects for data centres.

# 1 Introduction

The rapid growth of data-intensive applications, such as AI, machine learning, real-time analytics, and cloud services, drive robust computational infrastructures capable of processing vast data volumes at high speeds and efficiency[1–4]. The increasing demand for computational resources requires large-scale data centres to support extensive data storage, processing, and transmission, forming the foundation for modern data-driven services[5,6].

Large-scale data centres have evolved to meet these needs, serving as the backbone of cloud computing, big data analytics, and the Internet of Things. These data centres depend on the advanced infrastructure supporting high-capacity storage and high-performance computing systems, interconnected through sophisticated communication networks. As demand for computing power continues to surge, designing and optimizing data centres has become a major research focus, particularly in scaling performance while managing energy consumption and operational costs[7,8].

A recent and transformative development is the emergence of AI-DCIs, explicitly designed to meet the computational needs of AI and machine learning tasks. AI-driven DCIs require specialized hardware, such as graphical processing units and tensor processing units, to handle the parallel processing demands of AI models. In addition to hardware, AI-DCIs must be optimized for low-latency, high-bandwidth communications to meet the real-time data processing demands of AI applications while minimizing energy consumption and operational costs[9–11].

As these data centres scale to accommodate more complex AI workloads, the need for interconnecting systems through optical networks increases, creating challenges in data transmission speed, security, and reliability. AI-driven DCIs present unique challenges. Unlike traditional data centres designed for general-purpose computing, AI-driven DCIs require low-latency, high-bandwidth connections to support rapid data exchange between processing units[12,13]. Energy efficiency and scalability are critical as these systems grow in size and complexity[9]. Alongside these performance challenges, the growing threat of quantum computing to classical cryptographic protocols has raised serious concerns about data security[14,15]. Quantum computers, capable of solving some mathematical issues exponentially faster than classical computers, could easily break conventional encryption schemes, such as RSA and SHA. Particularly, recent breakthroughs in quantum computing have garnered significant attention, with both academic institutions and commercial entities achieving remarkable progress[7,16–20]. This urgently requires quantum-safe cryptographic solutions to secure communication channels in AI-DCIs.

Conventional solutions for interconnecting data centres, such as intensity-modulated direct detection (IM-DD) schemes and traditional intradyne coherent (IC) schemes, are unsuitable for meeting the stringent demands of AI-driven DCIs. While IM-DD systems are conventional and commonly deployed, they face challenges exceeding 1.6 Tb/s per channel, a key parameter for next-generation optical transmission in and between data centres[21,22]. While capable of high data transmission rates, traditional coherent systems face challenges in latency, complexity, and cost, making them impractical for large-scale AI applications[23,24]. These limitations underscore the need for advanced, secure, and scalable data transmission solutions that address the unique requirements of AI-driven data centres, including low latency, high bandwidth, data confidentiality, and cost efficiency.

We propose and demonstrate a quantum-secured data transmission architecture featuring DSP-Lite, designed to meet the stringent requirements of AI-driven DCIs. This architecture employs an SHC system for classical data transmission, simplifying the receiver structure while maintaining high sensitivity and robustness against phase noise, with a promising transmission capacity exceeding 1.6 Tb/s at low cost and low power consumption[25,26]. We employ QKD technology to withstand quantum computer attacks, ensuring that remote, legitimate users can exchange information-theoretically secure keys, impervious to the computational abilities of eavesdroppers[27,28]. To reduce interference between quantum and classical channels, we connect two remote users with multi-core fibres (MCFs), which provide minimal inter-core crosstalk and are compatible with existing single-mode fibre (SMF) infrastructures[29,30]. Furthermore, we develop an all-optical clock recovery through the remote delivery of a clock-embedded local oscillator (LO), which allows the QKD system to be seamlessly integrated into our quantum-secured network.

Our demonstration achieves a scalable transmission capacity of 2 Tb/s, with a quantum secret key rate of 229.2 kb/s via a 3.5-km MCF. Additionally, the QBER remains below 1.5%, even as transmission capacity increases, offering a cost-effective and power-efficient solution for high-performance data centre interconnections. Our work paves the way for the next generation of secure, scalable, and cost-efficient optical interconnects, protecting AI-driven data centres against quantum security threats while meeting the high demands of modern data-driven applications.

# 2   Results

## 2.1   Quantum-secured coherent communications for data centre interconnects (DCIs)

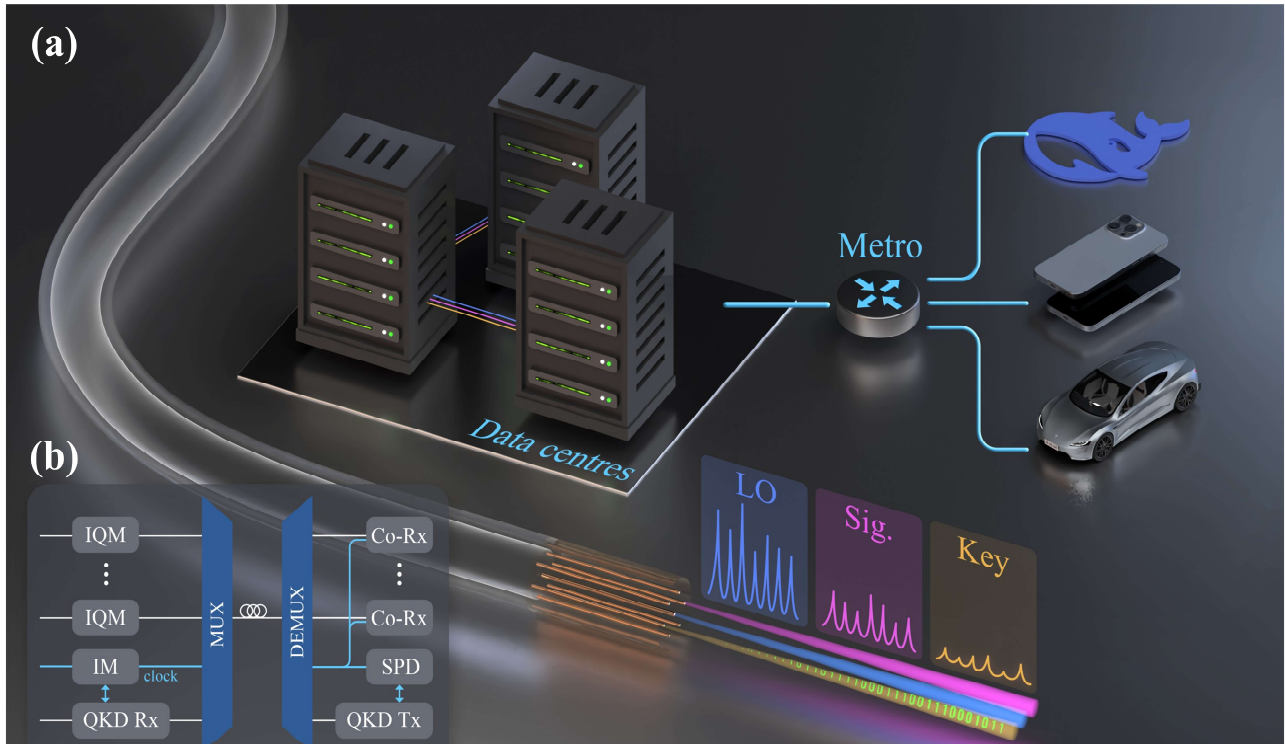

**Fig. 1: Concept and architecture of proposed quantum-secured networks in DCIs.** (a) Overview of the architecture. (b) The architecture of quantum-secured optical interconnect system. Classical data is transmitted using a SHC system, while the QKD technology provides security. In the quantum-secured optical interconnect system, data transmission relies on all-optical methods. The clock-loaded LO can synchronize the data transmission and QKD terminal, simplifying the architecture for secure data transmission.

Fig. 1(a) illustrates our proposed quantum-secured optical network for DCIs. Data centres are connected using MCFs. Classical data are transmitted using an SHC system and protected with QKD technology. Quantum-classical coexistence is achieved using SDM technology. This architecture, which integrates SDM and quantum-classical coexistence, enables secure data transmission at terabit-per-second (Tbit/s) rates, meeting the demands of future AI-driven applications such as smartphones, autonomous vehicles, and large language models. Our design prioritizes architectural simplicity and confidentiality to provide a cost-efficient solution for secure optical interconnect.

Fig. 1(b) presents the architecture of the quantum-secured optical interconnect system. For classical communication, we employ an SHC system consisting of intensity modulators (IM), in-phase quadrature modulators (IQM), coherent receivers (Co-Rx), and photodetectors (PD), while the IM and PD are responsible for generating and converting the synchronization signals. In our system, a homologous pilot tone is transmitted alongside a modulated signal through parallel spatial channels. The pilot tone acts as a remote LO and effectively reduces frequency offsets and phase noise during short-distance transmission[31,32]. Consequently, the system achieves high spectral efficiency (SE) and lower digital signal processing (DSP) consumption compared to traditional coherent detection. It also supports the use of low-cost, large-linewidth distributed feedback (DFB) lasers, significantly reducing commercial expenses[26,33–35]. The QKD system comprises QKD transmitters (QKD Tx) and QKD receivers (QKD Rx). For data transmission, modulated signals are encrypted using keys generated via QKD, with the secret key transmitted over a separate channel. The clock-embedded LO serves both to synchronize the transmitter and receiver for data transmission and as the synchronization signal for QKD.

Compared to conventional quantum optical networks, our scheme offers several distinctive benefits [36–39]. Primarily, the SHC system overcomes the limitations of traditional IM-DD schemes, which suffer from insufficient transmission rates and a limited interface speed of 200 Gb/s per channel. In contrast to standard IC schemes, our approach simplifies the system architecture and reduces DSP complexity. This architecture supports high-capacity data transmission while ensuring cost efficiency, structural simplicity, and compactness, thereby meeting the optical interconnect requirements for AI-driven data centres.

Secondly, we have integrated a QKD system to address the security concerns, which can be fully implemented using chip-based encoders and decoders[40–47]. The modulated carrier signal is then secured via an advanced encryption standard with a 256-bit key (AES-256) encryption, leveraging quantum-enhanced cryptographic protocols. The secret keys are transmitted through a dedicated core in the MCF, physically isolated from classical data channels. Furthermore, traditional coexistence schemes for quantum-classical systems typically require separate channels for QKD synchronization signals[29,38]. This cabling requirement for inter-DCI synchronization not only entails substantial energy consumption but also introduces infrastructural complexity. Using the proposed all-optical clock recovery, our architecture seamlessly integrates QKD with self-coherent detection systems. This approach achieves unprecedented stability in short-reach communications, enabling the future development of quantum-secured small form pluggable (SFP) optical transceivers.

## 2.2 Experiment setup

We conduct a coexistence experiment to validate our proposed architecture. Fig. 2 illustrates the schematic of the experimental setup, which comprises an SHC system and a polarization-encoding QKD system. The transmitter and receiver for both the SHC and QKD systems are connected via a seven-core fibre, with signals transmitted in opposite directions.

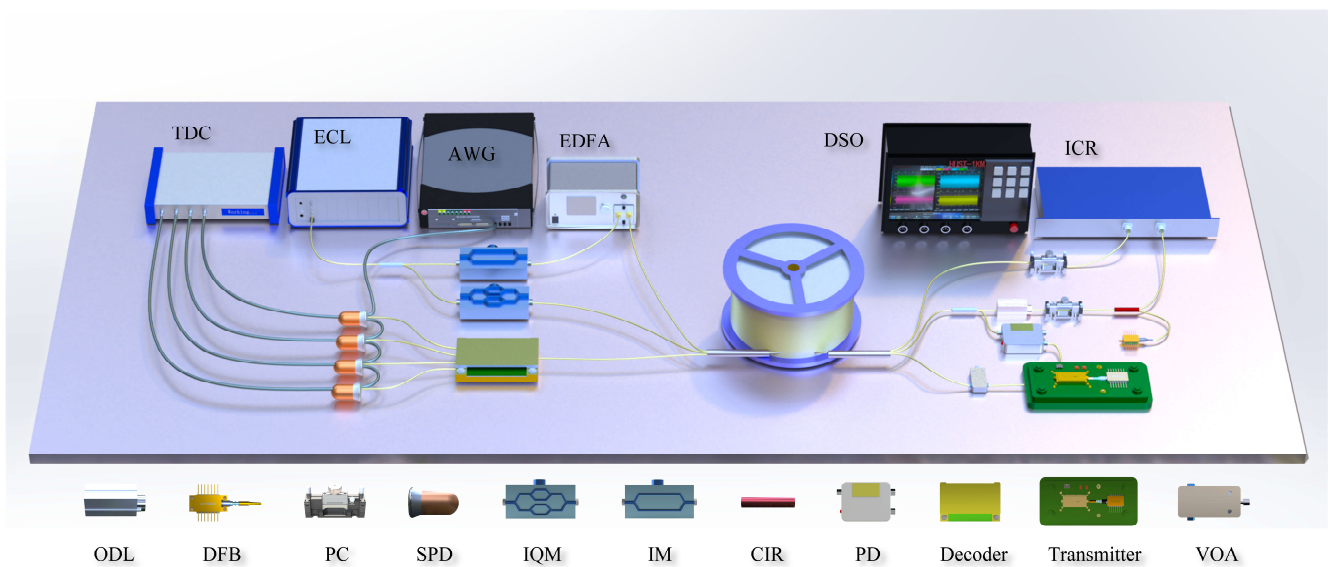

**Fig. 2: The schematic of the setup.** Quantum and classical communication are transmitted in opposite directions, with synchronization applied separately at both the transmitter and receiver sides. TDC time to digital converter, ECL external cavity laser, AWG arbitrary waveform generator, EDFA erbium-doped fibre amplifier, DSO digital storage oscilloscope, ICR integrated coherent receiver, ODL optical delay line, DFB distributed feedback laser, PC polarization controller, SPD single-photon detector, IQM in-phase quadrature modulators, IM intensity modulator,

CIR circulator, PD photodetector with a transimpedance amplifier, Decoder utilized for polarization decoding, sends four states into SPDs, Transmitter integrating a pulse laser with encoder, generates four states of polarization, VOA variable optical attenuator.

On the classical side, a 100 kHz-linewidth external cavity laser (ECL, MTP-1000) is employed as the light source, operating at 13.9 dBm. The light is subsequently divided into two paths using a polarization-maintaining power splitter. The upper path, carrying 90% of the power, is modulated by an intensity modulator (IM) and then amplified by an erbium-doped fibre amplifier (EDFA) to boost the LO. The lower path, also carrying 10% of the power, passes through a dual-parallel IQ modulator (DP-IQM) to generate the information-bearing optical signal. The modulated signal and the LO are combined and concurrently transmitted through two distinct cores of the seven-core fibre.

At the receiver, the fibre is connected to a homemade low-insertion-loss coupler (0.4–1.2 dB) that separates the transmitted light into two paths. The signal in the upper path is directly transmitted to the integrated coherent receiver (ICR), while the LO is further split using a 95/5 beam splitter. One of the split signals, carrying 95% of the power, is detected by a photodiode with a transimpedance amplifier (PD-w/TIA) and converted into electrical clock pulses to synchronize the quantum system. The lower path, carrying 5% of the power, passes through a polarization-maintaining circulator (PMC) to injection-lock the DFB laser. In our system, an optical delay line (ODL) is required for delay offset, and two polarization controllers (PCs) are needed for each channel to achieve polarization alignment.

For the quantum component, we construct a BB84 polarization-encoded QKD system with a commercial laser source operating at 100 MHz[48,49]. The laser is triggered by synchronizing signals from the photodetector (PD). The light pulses are coupled into an encoder consisting of an intensity modulator and a polarization modulator. The IM randomly modulates pulse intensities to generate decoy states, while the polarization modulator (PM) encodes quantum bits through polarization modulation. The quantum states are attenuated to single-photon level using a variable optical attenuator (VOA) before transmission through an MCF channel to the receiver. After passing through a narrowband filter for amplified spontaneous emission (ASE) noise suppression, the signals undergo polarization decoding before detection with a single-photon detector (SPD, model WT-SPD2000, Qasky Co. Ltd., 5.25% detection efficiency, 600 Hz dark count rate). Detection events are recorded by a time-to-digital converter (TDC, model quTAG100, qutools GmbH) and processed using a dedicated computer system.

All electronic signals are generated from a single high-speed AWG (Keysight, M9502A). This includes a signal at 100MHz to synchronize the decoder, the SPD array, and subsequently the time-to-digital converter (TDC), and a signal to trigger IM and DP-IQM.

## 2.3   Allocation strategy of SDM

Initially, we assess core allocation within the multi-core fibre (MCF) to ensure effective segregation of quantum and classical channels, as this impacts both classical data transmission and the SKR. We first analyze the impact of inter-core noise and incorporate inter-core spontaneous Raman scattering (ICSRS) noise into the background dark counts, obtaining an adjusted dark count value[50,51]:

$$p_R = \frac{p_{ICSRS}T_d}{2\varepsilon}\eta_R + p_{dc} \tag{1}$$

Where $\varepsilon$ is energy of a single photon and $T_d$ is gate width of detection, $\eta_R$ is the total transmission rate of a single photon, and $p_{dc}$ is the intrinsic background dark count rates.

We then conduct an experiment in which the classical signal, at various launch powers, is allocated to core 1, while dark count rates in cores 2, 3, and 4 are measured, respectively. The results, shown in Fig. 3, indicate that the dark count rate increases with higher launch powers and closely matches the simulated results obtained using Eq. (1). This further demonstrates that SDM technology enables efficient multiplexing within the C band, meeting the requirements of DCIs.

Based on these measurements, in the following experiment, core 4 is allocated to the LO and core

1 to the quantum channel, while the remaining five cores are designated for weaker signals.

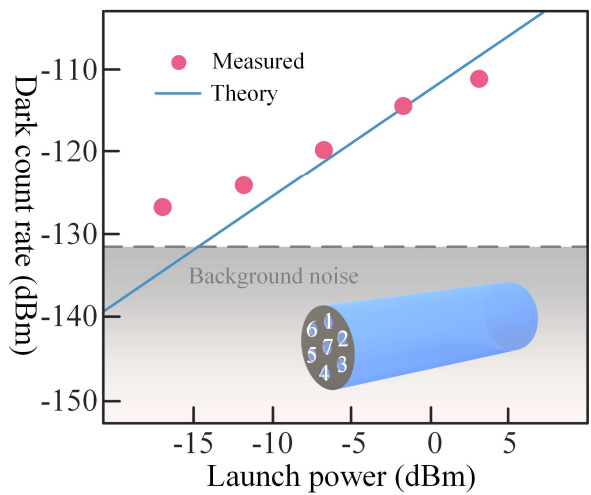

**Fig. 3: Dark count measurement and core selection.** Using a 3.5 km seven-core fibre, the measured dark count power was -112.1 dBm at a classical optical injection power of 2.04 dBm, consistent with the simulated result of -112.9 dBm under corresponding experimental parameters. The blue line denotes the simulated results where $T_d$=1ns, and $p_{dc}$ = 6×10$^{-6}$.

## 2.4   Testing of synchronization signals

We then evaluate the performance of our all-optical synchronization scheme in short-distance transmission. We directly measured the optical synchronization signal at the PD. For comparison, we also measured a signal from the radio frequency (RF) cable connection. The measurement results are presented in Fig. 4(a). The obtained FWHM for the cable-connection signal is 1.23 ± 0.07 ns, while for the optical synchronization signal, it is 0.97 ± 0.04 ns. These values are very close. The results indicate that, for short-distance data transmission, classical optical synchronization performs comparably to cable connection, accurately reflecting the arrival time.

To ensure that the low power and fluctuating LO can still be used effectively for coherent reception, the receiving end of the classical communication system has been optimized. This involves an injection-locked distributed feedback (IL-DFB) laser, which serves to mitigate the intensity fluctuations arising from the IM and regenerate the LO with high power and low noise for self-homodyne coherent detection (SHCD)[52]. In our experiment, the oscillating LO is restored at the receiving end, while the clock impairment is reduced from -12.5 dB to -23.1 dB by IL. (Fig. 4(b)). We choose a 3 GHz clock because it offers a better IL performance[53].

To illustrate the effectiveness of embedding the clock within the LO for coherent reception, the performance of various schemes is evaluated under different levels of received optical power, as depicted in Fig. 4(c). The optical clock scheme results in an approximately 3~4 dB penalty compared with the back-to-back (B2B) electrical synchronization by cable connection (E-SYNC), as well as the offline DSP of retiming. Our strategy achieves the same level of conventional cable connection, showing that it can be used in AI-DCIs.

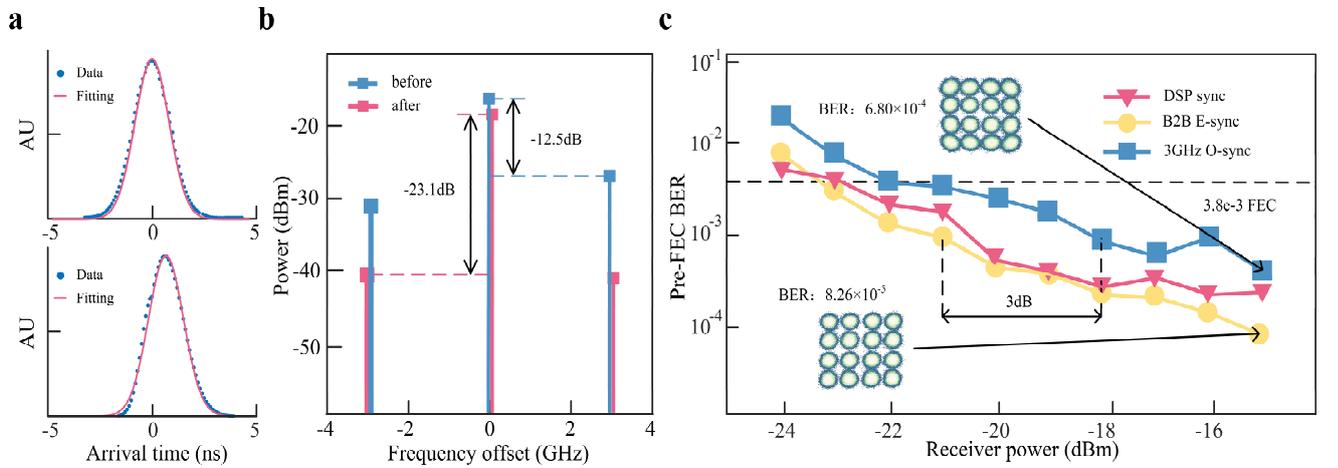

**Fig. 4: Classical communication segment of an all-optical network.** (a) Comparison of signal arrival time and FWHM between direct cable synchronization (top) and optical synchronization (bottom). The blue scatter points represent the measured signal sampling results, while the continuous red curve corresponds to the Gaussian fit. (b) In our strategy, the frequency offset is 0, and the IL-DFB laser reduces the clock interference on the LO. (c) Analysis of the cost of the optical clock synchronization. We illustrate the constellation diagrams with B2B synchronization and a 3 GHz clock oscillation.

## 2.5 Testing of classical and quantum rates

One of the main challenges for quantum communication is related to its performance at high classical power. To demonstrate the compatibility of our system, we further investigate the effects of various classical light power levels on QKD's SKR and QBER, illustrated in Fig. 5(a). Our system shows an average SKR of 229.2 kb/s ± 8.2 kb/s. The overall result is stable as the SKR decreases to 167.9 kb/s ± 7.6 kb/s even when the launch power surpasses 10 dBm. It demonstrates that the low inter-core crosstalk environment created by seven-core fibre enables our system to operate effectively in high-intensity light conditions, boasting significant advantages over other architectures[31,32]. Subsequently, to advance the practical deployment of a quantum-secured network, an experiment was carried out focusing on the transmission of encrypted data. Utilizing a 50 GBaud 16-QAM coherent transmission link, we demonstrate 400Gb/s data transmission rates for each core with quantum encryption. Fig. 5(b) illustrates that the SKR remains stable at 197.5 kb/s ± 6.0kb/s across various data transmission rates while also showing the QBER at 1.27% ± 0.02%, respectively. Incorporating a gradient-descent-based polarization feedback algorithm at the receiver, our system shows robustness.

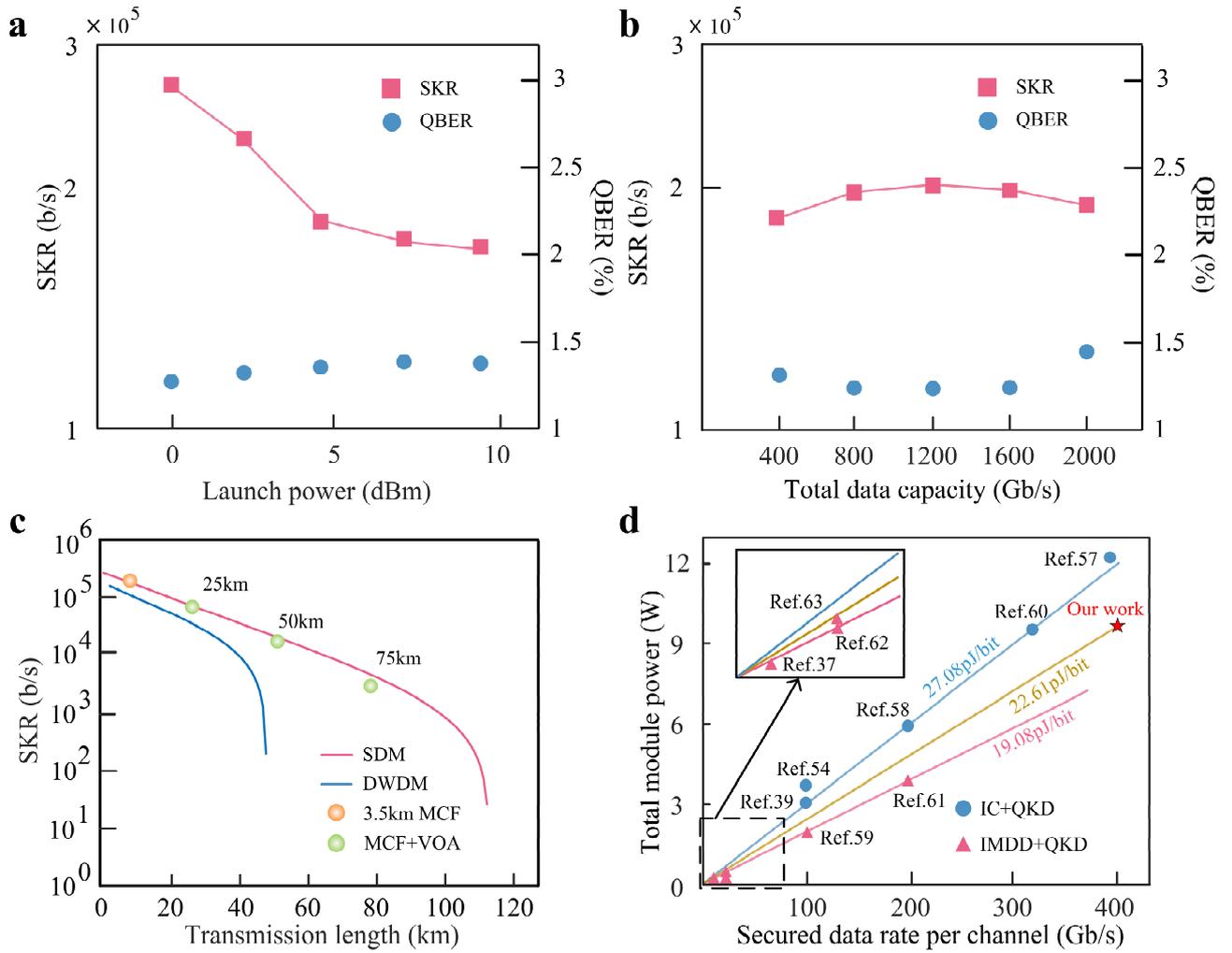

**Fig 5: Experimental measurement results and comparison with other schemes.** (a) SKR and QBER under various launch powers. (b) SKR and QBER under various launch powers. (c) Simulation and experimental signature rates under different transmission distances. The blue line represents the dense wavelength division multiplexing (DWDM) results, while the classical signal is 1310 nm. The red line shows the SDM strategy of 1550 nm, with a channel spacing of 100 GHz. The launch power of the classical signal is 10 dBm. (d) We compare the total energy consumption of different quantum-secured data transmission schemes, where the slope represents the energy consumption per bit.

The experimental results under different transmission distances are plotted in Fig. 5(c). The orange circle is the overall secret key obtained after propagation through the 3.5km MCF. Additionally, by applying an adjustable VOA to simulate transmit distance, we measure the SKR at distances of 50 km, 100 km, and 150 km. The experimental results demonstrate that we can generate substantial raw keys in MCFs of different lengths. The experimental results closely match the theoretical simulations. In conclusion, our strategy minimizes the ASE noise, yielding SKR comparable to that in dark environments. In contrast, the SKR generation in DWDM systems experiences substantial degradation at elevated optical power levels. All the calculations are conducted without filters.

We finally evaluate the total module power consumption to demonstrate the potential for reducing power consumption while enhancing coherent system performance. Based on 5 nm CMOS technology, we assessed the total module power consumption for a range of IC and IM-DD architectures to confirm the practicality and potential of SHC technology for internal data centre connectivity[54]. As Fig. 5(d) displays, compared to the IM-DD structure, our architecture achieves almost a two-fold increase in per-channel data rate. Meanwhile, it significantly reduces energy consumption relative to the IC structure while supporting larger data capacities per channel[37,39,55–62]. These results confirm that the proposed structure offers a compelling solution for simultaneously addressing transmission capacity, power efficiency, and security.

# 3    Conclusion and Discussion

We have demonstrated a quantum-secured data transmission architecture that addresses the critical challenges of AI-DCIs by integrating advanced optical communication technologies with quantum-safe protocols. Our demonstration achieves a 2 Tb/s classical data transmission alongside a 229.2 kb/s ± 8.2 kb/s quantum secret key rate over a 3.5-km MCF, with a QBER consistently below 1.5%. These results validate the feasibility of harmonizing ultra-high-capacity data transmission with provably secure QKD in a single infrastructure. This work bridges the gap between high-performance optical interconnects and quantum-safe infrastructure, offering a scalable blueprint for future AI-DCIs. By harmonizing cutting-edge photonics with quantum cryptography, we pave the way for secure, energy-efficient, and ultra-high-capacity networks capable of sustaining the exponential growth of data-driven technologies.

# 4 Methods

## 4.1 Experimental details

On the classical side, a continuous wave is generated by an ECL, with 90% of the power modulation data produced by an AWG, whereas the remaining light is utilized as a remote LO injected into another core. The input carrier is amplified by an erbium-doped fibre amplifier (EDFA) to a power of 0 dBm, and the modulated light is distributed to five cores through a coupler. At the receiving end, 95% of the LO is received by the PD-w/TIA, which then conveys the electrical signal to the QKD encoding end. The LO power is -11.2 dBm, which is subsequently amplified to 10 dBm through injection locking to another DFB laser with a similar structure. The power of the received signal light is -5.5 dBm, then sent to the ICR along with the LO. The receiver is powered by a direct current source (DC). The electrical signals from the received carrier and LO are processed through a direct current module and a balancer and captured by a high-performance oscilloscope (OSC). The OSC provides a sampling rate of 80 GSa/s, with the signals received by four channels, and the data are ultimately processed by a computer for DSP. At the receiving end of DSP, the sampled signals are first orthogonalized and normalized via the Gram–Schmidt algorithm to compensate for the influence of nonideal factors within the link. Subsequently, clock recovery is performed, followed by matched filtering of the signal with a root-raised cosine filter that has a roll-off factor of 0.05, and the signal is downsampled from 90 GSa/s to 50 GSa/s. Afterward, adaptive equalization is executed via the cascaded constant modulus algorithm (CMMA) and pilot-aided rotation phase recovery. The equalizer employs a 4×4 multiple-input output (MIMO) structure. Finally, symbol decisions are made, and the bit error rate (BER) is calculated to evaluate the system's performance.

On the quantum side, we perform a full optimization of the implementation parameters for the one decoy-state protocol. For example, in the scenario of 3.5 km, the intensities of the signal states and decoy state are chosen to be $\mu$=0.523 and $v$=0.131, respectively. The probabilities of sending out the signal state $\mu$ and choosing the bases Z are $P_\mu$=0.798 and $P_z$=0.898, respectively. At Bob's side, the probabilities of choosing the measurement basis in Z and X are permanently set to be equal due to the balanced MMI used in the decoder.

For each distance, the total length of periodic-correlation codes is 1024, and the ratio $M$ is set to 9:1. We continuously accumulate the detection numbers $n_z$ of approximately $10^8$ and perform the finite key analysis using

$$R \leq (s_{z,0}^L + s_{z,1}^L (1 - h(\phi_z^U)) - leak_{EC} - 6\log_2(19/\varepsilon_{sec}) - \log_2(2/\varepsilon_{cor}))/t \qquad (2)$$

where $t$ is the time duration of each acquisition, $s_{z,0}^L$ is the lower bound of the detector event received by Bob when Alice sends a vacuum state in Z basis, and $s_{z,1}^L$ is the lower bound of the detection event, given that Alice only sends a single-photon state in the Z-basis. $\phi_z^U$ is the upper bound of the phase error rate, $leak_{EC}$ is the number of bits consumed for error correction, and $\epsilon_{sec}$ and $\epsilon_{cor}$ are the parameters used to evaluate secrecy and correctness, respectively. The $h(x)$ denotes the Shannon binary information function

## 4.2    Noise Simulation and Calculation

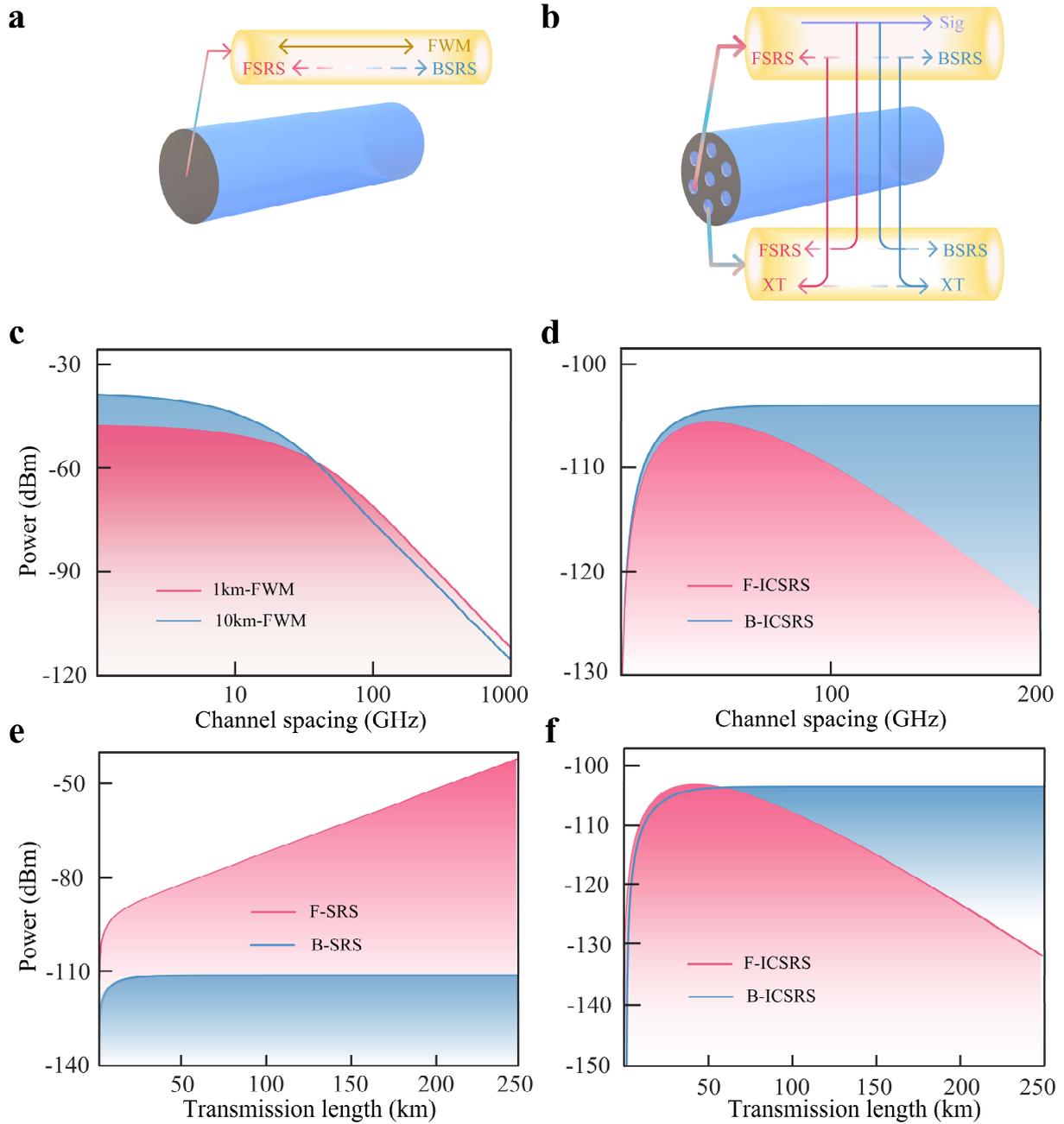

**Fig 6: Noise calculations in SMF and MCF in different frequency intervals and transmission distances.** (a) Noise model in SMF. (b) Noise model in a seven-core MCF. (c) Simulation of adjacent channel four-wave mixing (FWM) noise. We utilize two channels of equal power; $i$=$j$, and we set $P_i$=$P_j$=$P_k$=0dBm. We assume the classical signal is 1550 nm, then we change the spacing between the classical and quantum channel to calculate the FWM power. (d) Simulation of channel spacing forward inter-core spontaneous Raman scattering (F-ICSRS) and backward inter-core spontaneous Raman scattering (B-ICSRS) in seven-core fibre, six cores allocated to 0 dBm

classical signals. (e) Simulation of forward-SRS (F-SRS) and backward-SRS (B-SRS) in SSMF generated by two 0 dBm lasers. (f) Simulation of F-ICSRS and B-ICSRS under various transmission lengths in seven-core fibre, six cores for classical signals. The quantum signal is allocated to the remaining core.

We compared the predominant noise model in two transmission strategies (Fig. 6(a) and Fig. 6(b)). For DWDM, the noise is primarily Amplified Spontaneous Emission (ASE), dominated by SRS and FWM. When the classical signal wavelength is 1550 nm and the quantum signal is 1310 nm, we demonstrate the FWM noise introduced to QKD. The peak power of the FWM $P_{ijk}$ is given by:

$$P_{ijk}(z) = \frac{\eta D^2 \gamma^2 P_i P_j P_k e^{-\alpha z}}{9\alpha^2} [1 - e^{-\alpha z}]^2 \tag{3}$$

where $P_i$, $P_j$, and $P_k$ denote the power levels of the three input channels, with $\alpha$ representing the fibre attenuation coefficient, $z$ representing the propagation distance, and $\gamma$ is the fibre nonlinearity parameter. $D$ is the FWM degeneracy factor. Conversely, when only two optical channels are present and share the same frequency ($i=j$, $f_i=f_j$), the FWM process degenerates, and $D=3$. The FWM efficiency is:

$$\eta = \frac{\alpha^2}{\alpha^2 + \beta^2} \{1 + \frac{4e^{-\alpha z} \sin^2(\Delta\beta z / 2)}{[1 - e^{-\alpha z}]^2}\} \tag{4}$$

The $\beta$ represents the propagation constant of the various input channels ($i$, $j$ and $k$) and the resulting mixing product ($ijk$), while $\Delta\beta$ is the phase matching factor.

In the SDM strategy, we also adjusted the total dark count rate by the ICSRS model:

$$\begin{aligned} P_{total} &= P_{dark} + P_{ICXT \to FSRS} + P_{FSRS \to ICXT} + P_{ICXT \to BSRS} + P_{BSRS \to ICXT} \\ &= P_{dark} + P_{F-ICSRS} + P_{B-ICSRS} \\ &= P_{dark} + P_{ICSRS} \end{aligned} \tag{5}$$

where $P_{dark}$ represents the background dark counts, $P_{FSRS \to ICXT}$ is the F-SRS generated in the quantum channel by crosstalk from the classical signal, $P_{FSRS \to ICXT}$ is the crosstalk in the quantum channel caused by forward SRS from the classical signal, $P_{ICXT \rightarrow BSRS}$ and $P_{BSRS \rightarrow ICXT}$ similarly accounts for the B-SRS between cores (Fig. 6(d)). $P_{F-ICSRS}$ represents the power of total forward inter-core Raman scattering, while $P_{B-ICSRS}$ denotes the backward inter-core Raman scattering.

The expressions are as follows:

$$P_{F \rightarrow ICSRS} = P_0 \eta_{ij} \exp(-\alpha_q L) \{ \frac{\exp[(\alpha_q - \alpha_c)L] - 1}{\alpha_q - \alpha_c} - \frac{\exp[(\alpha_q - \alpha_c - 2h_{ij})L] - 1}{\alpha_q - \alpha_c - 2h_{ij}} \} \qquad (6)$$

$$P_{B \rightarrow ICSRS} = P_0 \eta_{ij} \{ \frac{\exp[-(\alpha_q + \alpha_c + 2h_{ij})L] - 1}{\alpha_q + \alpha_c + 2h_{ij}} - \frac{\exp[-(\alpha_q + \alpha_c)L] - 1}{\alpha_q + \alpha_c} \} \qquad (7)$$

$P_0$ is the power of the classical signal, which is set to 0 dBm at 1550 nm. $\eta_{ij}$=6.9×10⁻⁸ nm⁻¹km⁻¹ is the Raman efficiency, and $\alpha_q$=0.046 km⁻¹, $\alpha_c$=0.0471 km⁻¹is the attenuation coefficient for the quantum and classical channels[63]. The coupling coefficient, $h_{ij}$, can be ascertained by simulating the transmission distance and crosstalk[64]:

$$ICXT = h_{ij}L \qquad (8)$$

By fitting multiple sets of crosstalk and transmission distance data, $h_{ij}$ is set to 7×10⁻⁷.

We first present the power of FWM and ICSRS as a function of varying channel spacing (Fig. 6(c) and Fig. 6(d)). For simplicity, we only plot the peak power of the beam modulation, ignoring the oscillatory nature of the beam modulation efficiency. It demonstrates that the calculated FWM is negligible at channel spacing over 1000 GHz. Consequently, we compare the SRS and ICSRS under various transmission lengths. The ICSRS can remain at levels close to dark counts (Fig. 6(e)). However, the B-SRS is observed around -100 dBm, but F-SRS initiates at -110 dBm and increases with transmission distance, significantly surpassing the dark count level, shown in Fig. 6(f). Consequently, it is generally believed that we can only employ a combination of the O+C bands to facilitate the coexistence of quantum and classical signals. However, the attenuation coefficient of the O band is significantly higher than that of the C bands (0.093 km$^{-1}$ compared to 0.046 km$^{-1}$), presenting a disadvantage for transmission.

## 4.3    Energy Consumption Calculation and Comparison

In contrast, the consumption of QKD is benchmarked against the one-time pad standard. The energy consumption of the Tx per information bit $E_{tx}$ is given by[65,66]:

$$E_{tx} = \eta^{-1}(E_{laser} + E_{mod} + E_{enc} + E_{dac}) \tag{9}$$

where $E_{laser}$ is the laser energy consumption per information bit, $E_{mod}$ is the modulator energy per information bit, including encryption using AES-256. $E_{enc}$ is the encoder energy per information bit for QKD. $E_{dac}$ is the digital-to-analog converter (DAC) energy consumption per information bit and $\eta$ is the transceiver module power conversion efficiency. Similarly, the energy consumption of the Rx per information bit $E_{rx}$ is given by:

$$E_{rx} = \eta^{-1}(E_{PD} + E_{ADC} + E_{DFB} + E_{DSP} + E_{sift}) \tag{10}$$

where $E_{PD}$, $E_{ADC}$, $E_{DFB}$, and $E_{DSP}$ are the energy consumption per information bit for the PD, analog-to-digital (ADC), DFB, DSP, and sifting. In our experiment, post-processing is confined to the sifting phase, so we exclusively account for the energy consumption during sifting. The power consumption of DAC and ADC methods was determined based on the sampling rates and resolutions specific to each scheme.

## Acknowledgments

This work was supported by the Innovation Program for Quantum Science and Technology (No. 2021ZD0300701), Hubei Optical Fundamental Research Centre, the National Natural Science Foundation of China (No. 62205114 and No. 62225110), and the Major Program (JD) of Hubei Province (No. 2023BAA001-1).

## Author contributions

The concept of this work was conceived by S.Y. and K.W. The experiments were performed by X.J. and W.H., with the assistance of Y.L., J.C., and M.Z. The results were analysed by X.J. and W.H. X.J., S.Y., and K.W. wrote the first draft of the manuscript. S.W. and M.T. revised the manuscript. All the authors participated in the revision of the manuscript. The project was supervised by S.Y., K.W., S.W., and M.T.

## Competing interests

The authors declare that they have no competing interests.